# A classical flow instability and its connection to gaseous galactic disk hydrodynamics

Georgios H. Vatistas

Department of Mechanical and Industrial Engineering
Concordia University
Montreal Canada

Abstract
Water vortices are known to develop polygon structures inside their cores. The apexes of the polygonal manifestations are the result of satellite vortices attached to the parent vortex. Exploiting the analogy between the shallow water hydraulics and the two-dimensional compressible gas flows we arrive to the following line of reasoning concerning the spiral galactic structure. The backward free surface fore ripples of the advancing pressure disturbance in the hydraulic system should appear as high-density spirals in the gaseous galactic disk. Since the number of galactic "arms" is equal to the number of satellite eddies in the parent vortex, spiral galaxies with say two arms will possess two nuclei. The last appears to be in complete agreement with the latest observations regarding M 31 and Mrk 315 galaxies. If the present hypothesis is accurate then a closer look at the central regions of multi-tentacle spiral galaxies should also reveal that these have multiple nuclei.

Introduction

Whirlpools produced in simple confinements have often been employed to elaborate on some fundamental properties of rotating flows. Isaac Newton demonstrated the importance of the centrifugal force through a revolving "bucket" of water. Stationary dishes with revolving devices at the bottom have been utilized by several investigators to study the salient properties of a variety of complex natural phenomena. For example Vettin (1857) used a rotating disk with a centrally located cylindrical cup filled with ice to simulate the earth's polar circulation. Fridman et al. (1985) examined the role of centrifugal instability in the development of the spiral structure of galaxies. Their experimental apparatus featured a stationary dish with an independently rotating ring/conical cup-like device at the bottom of the container.

The general theme of the present paper is an old one first studied theoretically by Lord Kelvin in 1880. His work on the vibrations of a columnar vortex assumed perturbations in the three space dimensions with a constant interface in the axial direction. Our involvement with the phenomenon began while investigating particle concentrations in liquid vortices, produced within a stationary cylindrical container, using revolving disks and cylinders Vatistas (1989). During a detailed mapping of the "steady" free-surface elevation several problems were encountered that were later attributed to the dynamic nature of the core. In slender tall vortices surface folds at higher elevations interfered visually with the shapes created near the rotating disk surface. In order to avoid this obstruction of the view, we lowered the liquid height. Then the attractive polygonal shaped imprints on the disk surface came into full view. The experience, among some other captivating related phenomena, was reported in a series of articles, Vatistas (1990), Vatistas et al. (1992), (1994) , and (2001).

Recently professor Bohr and his research team confirmed independently (see Jansson et al. 2006) our 16 years earlier observations and description of the event. Although the phenomenon described in the previously mentioned paper is not new, the current relevance of the subject matter to various scientific disciplines triggered a worldwide publicity.

The fundamental nature of the phenomenon

Past experimental studies, Vatistas (1990), Vatistas et al. (1992), and (2001), have shown that under prevailing conditions, waves on the free surface of a liquid vortex exhibit the fundamental characteristics of Kelvin's equilibrium patterns. The rotary motion imparted to the fluid by the disk, see figure 1, generates a centrifugal force field, which pushes the liquid towards the container's wall. The receding liquid exposed part of the surface of the disk to air whereby, the line of intersection between the surfaces of the solid disk, the liquid, and air outlines the core shape. In order to bring the patterns into relief, the liquid was colored with a blue water-soluble dye. For very low rotational disk speeds it is expected that the liquid (water) vortex core remains circular (mode $n = 0$). Increasing its rotation, the vortex will transfer into another state characterized by a precessing circular core ($n = 1$). A further increase of disk speed ($\omega_d$) yielded progressively cores with elliptical ($n = 2$), triangular ($n = 3$), square ($n = 4$), pentagonal ($n = 5$), and hexagonal ($n = 6$) cross-sections. No heptagonal shape was able to form. Since the interval of endurance of the stationary states decreases with $n$, if $n = 7$ exists in theory it must be critically stable. As the disk speed increases well beyond $n = 6$ a continuous amplification of dynamical noise eventually wipes out the sharp spectral peaks.

The equilibrium polygons were found to be exceptionally stable. When disturbed by a momentarily applied external disturbance to the flow, the patterns reemerged after a short period of time in their original form. Both quasi-static and sudden increase to the final disk speed produced the same equilibrium pattern. The latter indicated that the phenomenon is not particularly sensitive to initial conditions.

Between neighboring states mixed-mode time dependent equilibria were found to exist. For example between the $n = 3$ and $n = 4$ band, the core consists from the superposition of both waves. Because the two have different phase speeds the core appeared to be non-stationary.

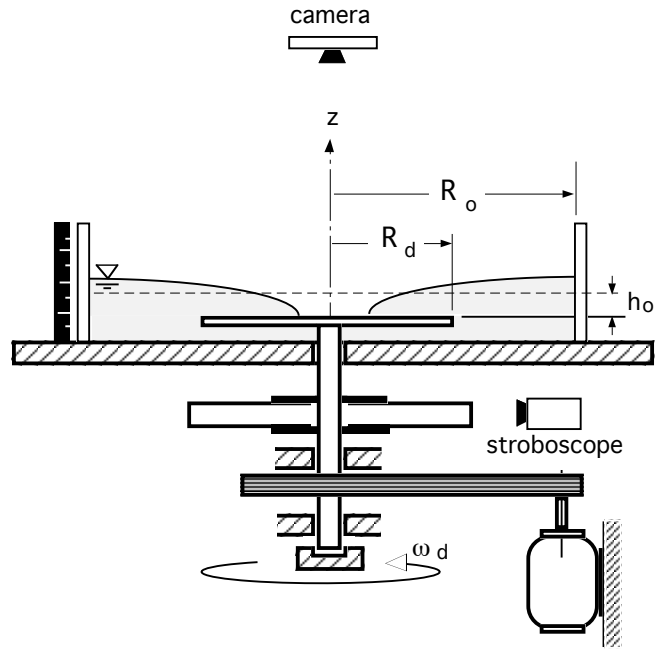

Figure 1: Schematic of the experimental apparatus. The device consists of a stationary plexiglas cylindrical container, with an aluminum disk rotating in the counter - clockwise direction at the bottom of the container. The shaft is driven, through a belt pulley assembly by a variable speed motor. A relatively large metallic flywheel provides additional inertia to keep the rotation constant. The electric motor develops speeds ranging from 0 to 1500 rpm. The initial water level was measured using a ruler attached to the side of the tank. A photo reflective digital meter determines the disk angular velocity. A stroboscope was used to verify the values of the angular velocity of the disk.

In our early experiments, visual inspection revealed that the patterns were present even if the core was flooded. Recent higher fidelity images shown in figure 2 (a) and (b) confirm the old observation. Consider for example figure (a) and (b). In the proximity of the core, a "circular" dry spot exists. However the region inhabited by the apexes of the polygon is flooded with water. The light coloring in each "corner" indicates a free surface depression. The last is the consequence of the centrifugal force developed by each of the satellite vortices. Hence, these polygonal shapes can be either viewed as instability waves inside the vortex, or as rotating ensembles of satellite vortices attached to the parent vortex.

Two types of free surface sets of ripples are present in all the photographs, and particularly in (c). One kind occurs in front, while the other trails behind the advancing pattern. The moving polygon represents a free surface pressure disturbance moving in a circular orbit at supercritical speeds. The last gives rise to backward spirals that are the polar equivalent of Kelvin's ship fore waves, Kelvin (1987) and Stoker (1992). Behind the pattern another group of ripples are the accompanied Kelvin's ship aft (or transverse) waves. A turbulent patch clearly visible in (b) appears to be similar to Emmons turbulent spot developing on a flat plate during the transition from laminar to turbulent boundary layer, visualized by Chantwell et al. (1978).

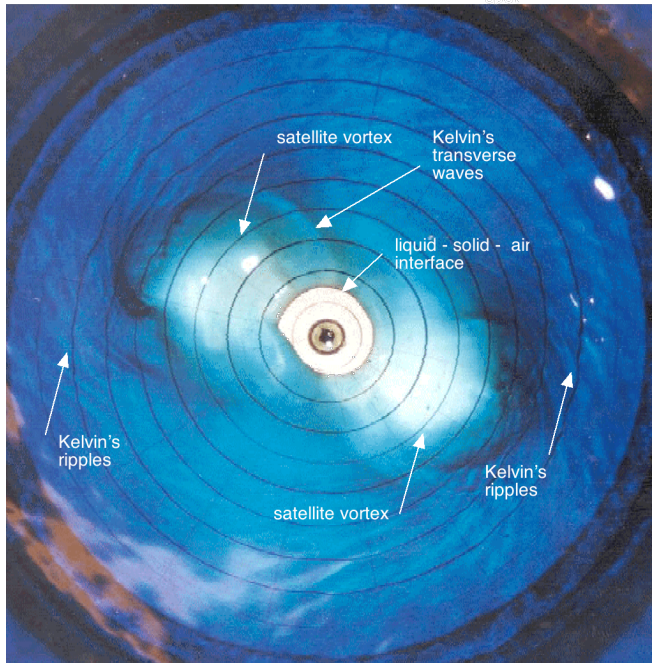

(a)

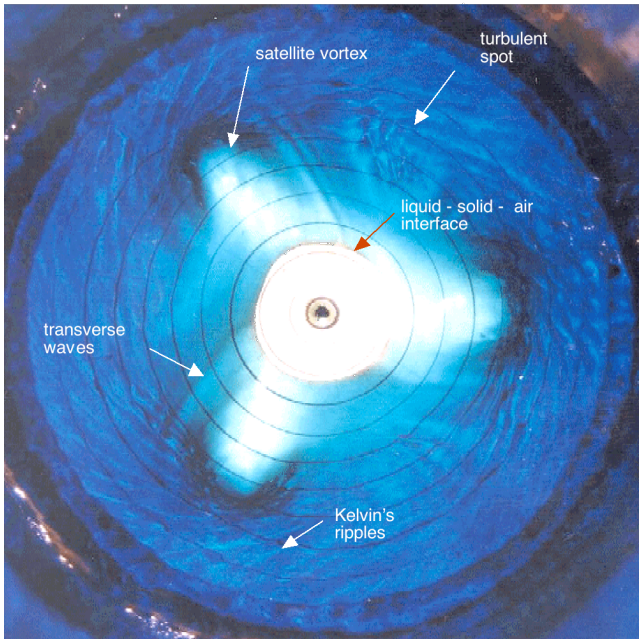

(b)

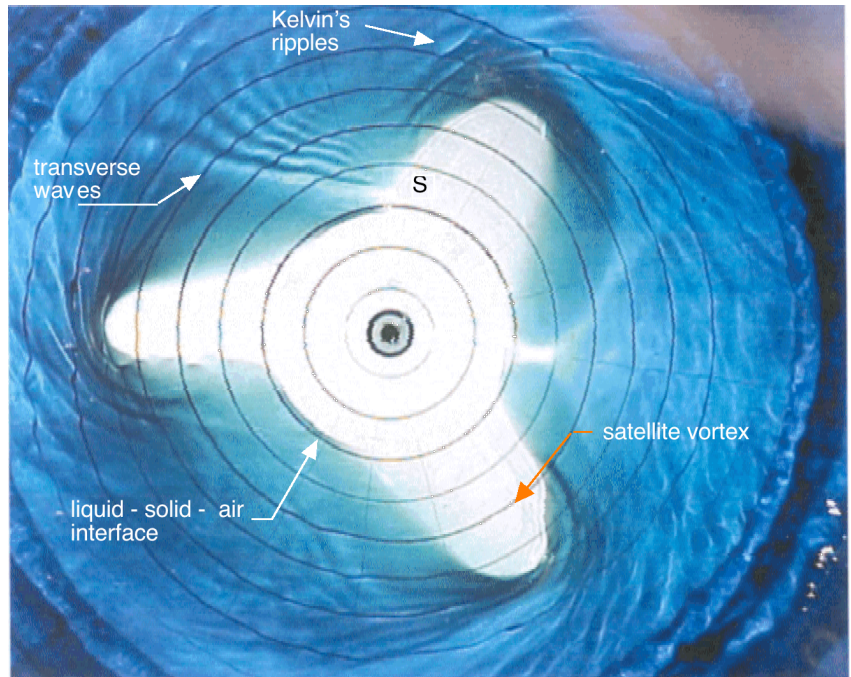

(c)

FIGURE 2: The centrifugal force imparted to the fluid by the disk compels the liquid outwards. The retreating liquid exposes the structure of the core. Every vortex with an interface develops a free-surface depression near the axis of rotation. In the corners of the polygonal core shapes satellite vortices exist. Each satellite vortex possesses a dimple. The last is very evident in (a) and (b) where the faded color polygon tips indicating a free surface depression due to the satellite vortex. Since the parent vortex is stronger, the central portion is dry. When the rotation intensifies both parent and satellite vortices become stronger. The receding water develops lobes in the place where the secondary vortices existed, and thus forms the propeller-like dry pattern shown in (c). A turbulent patch similar to Emmons in boundary layers appears in (b).

The link of the phenomenon to galactic vortices

The study of physical phenomena by analogy is a common and useful method of scientific inquiry that has been applied in many physical areas. Two systems, that may not necessarily physically resemble each other, are considered to be analogous, if both are described by the same set of the dimensionless equations. One of the most well known examples of such systems is the shallow water hydraulics and the two-dimensional compressible gas flows. Landau & Lifshitz (1987). Although the topology for this analogy is not, in the formal sense complete, Thompson (1972), shallow water experiments have been used in the past with success to gain a qualitative understanding of certain compressible gas flow properties. In addition, the most salient characteristics do also come into view even if the problem is not purely two-dimensional and inviscid, the corresponding gas does not have a specific heat capacity ratio of two, and/or the water level is low.

During our tests we noted the similarity the laboratory eddy under consideration with galactic images. The "arms" were clearly visible when milk was poured into the core cavity. Many astrophysical problems have been viewed in the past as paradigms of fluid motion, see for example Fridman et al. (1985) and Lin & Roberts (1981). But, how could an insignificant (in the grand scheme of things) laboratory experiment like ours reveal important characteristics of a grand celestial design? According to Fridman et al. (1985): *" … the evolution of perturbations (instabilities) in the gaseous galactic disk, resulting in the formation of spiral density waves, is determined essentially by hydrodynamics, …"*. Therefore, by analogy the hydraulic simulations can indeed uncover some of the most essential characteristics of the grand design.

In 1993 Lauer et al. reported that Andromeda galaxy (M31) possesses a double nucleus. Lauer and colleagues attributed the reason of the manifestation a cataclysmic collision of two venerable galaxies. Based on our experience with the hydraulic analogue we proposed, Vatistas (1993), that vortex instability may also be a reasonable alternative cause for this effect. Recently another galaxy Mrk (Markarian) 315 was also found to possess two nuclei.

Each satellite vortex generates a backward spiral. These free surface ripples will appear as high-density spirals in the gas dynamic analogous system, or the arms. Therefore, the number of arms present in a galaxy should be equal to the number of satellite vortices or nuclei. For galaxies with $n = 1$ there should be a single core that is precessing and thus the galaxy should possess one backward spiral. Galaxies with $n = 2$ must have two nuclei and hence develop two arms. There is a score of galaxies possessing the last attribute. Galaxies known to have 3 and 4 arms are for example the M39 and NGC 1232 respectively. Andromeda (M31) is now identified to be a galaxy with two cores. Previously it was well known that Andromeda possessed two spiral arms. Mrk 315 has been recently found to have two nuclei. Consistent with the conjecture, a new photometric and spectroscopic observations of Mrk 315 by Ciroi et al, (2005) revealed the presence of two faint spiral arms. If the present hypothesis is accurate then a closer look at the central regions of multi-armed spiral galaxies, such as for example the NGC 613 should also reveal multi-nuclei structure (note that the galactic disk is inclined by $32^o$ from the vertical axis of the image plane).

The current discovery of the double ring in the dwarf galaxy Mrk 409 by Git et al. (2003) is the consequence of waves developing in the axial direction. The vortex in the radial-azimuthal ($r$-$\theta$) plane is of the mode $n = 0$. The vortex however develops also instability waves in the axial ($z$) direction. Depending on the axial wave number the last will provide the breathing mode (sausage-like free surface corrugations) whereby the nucleus will emit periodically gaseous compression waves that will be seen from afar as multiple high-density rings.

Conclusions

This paper described some of the physical aspects of polygonal vortex core formations and explored their relevance to the gaseous galactic disk hydrodynamics. The multi nuclei manifestations in spiral galaxies were attributed to the presently dealt with instability. It is indeed astonishing to find that a lesser experiment like the present could unravel some of the secrets of phenomena occurring in such a grand scale.